\documentclass[aps,prstper,preprint,groupedaddress,amsmath,amssymb]{revtex4}

\usepackage[normalem]{ulem}	

\usepackage{graphicx}

\usepackage{dcolumn}
\usepackage{bm}
\usepackage{verbatim} 
\usepackage{units}
\usepackage{tipa}
\usepackage{xcolor}
\usepackage{booktabs}
\newcommand{\ra}[1]{\renewcommand{\arraystretch}{#1}}

\begin{document}


\title{Transforming a 4$^{th}$ year Modern Optics Course Using a Deliberate Practice Framework}
\author{David J. Jones}

\email{djjones@physics.ubc.ca}
\author{Kirk W. Madison}
\affiliation{Department of Physics and Astronomy, \\University of British Columbia, Vancouver, Canada}

\author{Carl E. Wieman}
\email{cwieman@stanford.edu}
\affiliation{Department of Physics and \\ Graduate School of Education,\\ Stanford University, Stanford, CA}



\date{\today}

\begin{abstract}
We present a study of active learning pedagogies in an upper division physics course.  This work was guided by the principle of Òdeliberate practiceÓ for the development of expertise, and this principle was used in the design of the materials and the orchestration of the classroom activities of the students.  We present our process for efficiently converting a traditional lecture course based on instructor notes into activities for such a course with active learning methods. Ninety percent of the same material was covered and scores on common exam problems showed a 15\% improvement with an effect size greater than $1$ after the transformation. We observe that the improvement and the associated effect size is sustained after handing off the materials to a second instructor.  Because the improvement on exam questions was independent of specific problem topics and because the material tested was so mathematically advanced and broad (including linear algebra, Fourier Transforms, partial differential equations, vector calculus), we expect the transformation process could be applied to most upper division physics courses having a similar mathematical base.
\end{abstract}


\maketitle


\section{Introduction} \label{sec:intro}

There is a very extensive literature showing that a variety of active learning strategies lead to better gains than traditional lectures on concept inventories and other similar measures of conceptual mastery of introductory physics. ÊA review across the sciences was commissioned by the National Research Council \cite{NRC_DBER} and, in a specific paper within this report, Docktor and Mestre specifically synthesized the results of physics education research (PER) to date \cite{Docktor}. ÊAs these results are so extensive, it is often assumed that these teaching strategies have been shown to work equivalently in upper-division physics courses. To our knowledge, this assumption remains untested. We believe that there are several arguments as to why such teaching strategies that have been successful in lower division undergraduate courses will not necessarily lead to the desired (and similar) improvements in student mastery within upper-division courses, including more sophisticated mathematics, more difficult/complex concepts, and a student population that is inherently filtered to be highly motivated/interested in physics. We have heard these arguments repeatedly expressed by our non-PER physics colleagues. Here we present our view of current research on the effectiveness of various instructional strategies in upper-division physics courses and then present arguments for why somewhat different teaching strategies from those used in introductory physics courses may be necessary for upper-division courses. ÊFinally, we will offer an extension to the conventional theoretical frameworks of Òactive learningÓ or Òinteractive engagementÓ that provides guidance for how to extend the successes of PER into the different contexts of upper-division physics courses. Ê We then present an example of the transformation of a specialized upper-division physics course using this framework, and the resulting quantitative Êimprovements in student learning that are obtained. 

\subsection{Existing PER literature on the effectiveness of upper-division course transformations}

Very few published papers report comparisons of learning between two different teaching methods in an upper-division physics course.  There have been a number of papers on the transformation of upper-division courses that use novel methods or curriculum, but they have very limited measures of learning and, with one exception, no such comparative data indicating the superiority of one teaching method over another. 
For example, some PER work on upper division courses has focused on the characterization of persistent difficulties with fundamental concepts \cite{NeedForTutorialApproach} and mathematics \cite{pepper_observations_2012} that hinder meaningful learning of advanced topics.  This work motivated the creation and use of tutorials where the instructor has small groups of students work collaboratively on worksheets that guide them through proper reasoning to overcome misconceptions and gain physical insight or that simply provide supervised practice of key mathematical tools \cite{NeedForTutorialApproach,DevelopingTutorials}.  Other upper division PER work has investigated the improvement of learning outcomes and student attitudes after a careful re-structuring of large parts of the curriculum or of the learning goals of courses \cite{ParadigmsInPhysics,Wagner_2012,PhysRevSTPER.8.020107}.  Two examples of course transformations are the work to implement active learning in an intermediate optics \emph{laboratory} at Indiana University-Purdue University Indianapolis \cite{ActiveLearningIntermediateOptics}, and the work at Kansas State University to implement the studio format for instruction in an upper-division optics lecture course \cite{StudioOptics}.  Limitations of both studies are that the sample sizes were very small (7 and 15 respectively), and each study reported the results from a single year with no comparison made with their pre-reform counterparts.

The lack of comparison with pre-reform courses is a serious limitation since without comparison, it is difficult to assess the actual impact of course reform on student learning, and as a result the efficacy of these methods in upper-division courses remains an open question. The sole exception is the work from the University of Colorado by Pollock and coworkers.  They have carried out a 4-year study of an upper-division E\&M course that was transformed using the principle of active engagement \cite{PhysRevSTPER.8.020107}.  That study reports the results of both conceptual tests as well as traditional exam questions for both pre and post-reform student populations.  The results show that these interventions resulted in higher learning gains compared to those in the traditionally taught courses in the areas of student's conceptual understanding and ability to articulate their reasoning about a problem; however, the comparison on basic correctness of traditional exam problems shows no statistically significant difference between the transformed and untransformed versions.

This difference in the results on conceptual tests compared to those involving detailed quantitative problems is consistent with the PER results from introducing new teaching methods into introductory courses. While there are many reports of improved results on conceptual inventories and targeted conceptual tests, the results on standard quantitative problems are often unchanged \cite{Docktor}. This raises serious questions as to the efficacy of such teaching methods in upper-division courses, particularly the most advanced ones.  At that level one is working with material where a practicing physicist would see little distinction between ``conceptual'' and ``qualitative problem solving'' mastery.  The ability to translate conceptual understanding into mathematical equations and quantitative calculations, as well as the inverse, is the basic expertise that a physicist uses on a routine ongoing basis.  A primary learning goal is for students to gain that expertise.  Given that these active learning methods have not shown improvements in students skill at doing quantitative calculations, it raises serious questions as to whether these methods will lead to improved educational outcomes in upper-division courses, where such skill is an essential goal of the instruction.

\subsection{Current theoretical frameworks in PER}

As reviewed by Docktor and Mestre, current instructional theoretical frameworks used in PER are based on constructivism and involve various ways to have the students be more mentally active than when listening to a lecture.  These strategies involve a variety of different types of activities including: ``...assigning open-ended tasks (e.g., problems) that are authentic and challenging...Ó, ``...opportunities to work collaboratively with their peers...'', ``...providing appropriate scaffolding for activities...'',  ``...exploration, concept introduction, and concept application (Karplus learning cycles)...'' \cite{Docktor}. There are ``concept tests'' during lecture and ``peer instruction'' \cite{mazur_peer_1997}.  Hake defines the whole category as {\it interactive engagement}: ÒÉmethods as those designed at least in part to promote conceptual understanding through interactive engagement of students in heads-on (always) and hands-on (usually) activities which yield immediate feedback through discussion with peers and/or instructors.'' \cite{hake_interactive-engagement_1998}.

In our view, approaches that do not directly address the nature of the questions students should be answering and discussing are under-specified. In particular, if no guidance is given as to the discipline-specific cognitive activities that are needed, the desired mastery may not be obtained.  A lack of explicitness as to the underlying principle behind an instructional design may limit the fidelity of its use in instruction. We speculate that many of the choices made by PER researchers and instructors as to the specifics of the active learning tasks they implement are implicitly driven by their educational goals, as defined by what they measure.  When they are using concept inventories that measure students' ability to correctly apply the relevant concepts in simple systems, the instructor will introduce methods like concept tests into lectures, tutorials and recitations, where students answer questions on the basic concepts and their application in a variety of simple situations.  It is not ``teaching to the test'' in the sense that it is asking students the same questions as on the test, but it is ``teaching to the test'' in the sense that it is giving students practice in the same kind of conceptual reasoning and near transfer that is tested by the concept inventories.  That is entirely appropriate as such reasoning is (and in our opinion should be) an important educational goal of such courses. However, it is not focusing on practice and feedback on improving quantitative problem solving.  

\subsection{Deliberate practice as a theoretical framework for PER}

Here we use the concept of ``deliberate practice'' from educational psychology, with some refinement, as a theoretical framework for giving the specificity needed for designing ``active learning'' tasks that will achieve specific educational goals that may be different from those already studied.    Deliberate practice \cite{ericsson_role_1993}  is the intense explicit practice with timely guiding feedback of the specific skills that make up expert performance in a discipline.  Our refinement is to focus on the specific cognitive tasks that make up the desired expertise in physics and the selection of tasks that provide students with practice in carrying out those cognitive processes, at the level which provides them with appropriate challenge \cite{deslauriers_improved_2011}. This is identifying the specific components of the thinking that should be practiced. Our ``deliberate practice'' framework is a way to dissect the expert skills used in an area of physics in order to define tasks that practice them, similar to the way a good tennis coach dissects the skills needed to be a good tennis player and creates tasks that practice the specific skills.

Determining those cognitive tasks requires ``cognitive task analysis'' of the relevant expert thinking.  We have found it particularly useful to focus on when and where experts make decisions in working through a problem in the area of interest.  Tasks are then designed for the students tasks that embody the thinking involved in those decisions. We are making explicit, and hence more easily transferrable, basic ideas that have been used extensively, but implicitly, in much of the existing PER work.  Concept inventories are largely based on finding specific conceptual reasoning processes that are completely routine to experts/physicists but on which students do badly.  Those active learning methods that show improved gains are then giving students practice and feedback on doing the type of conceptual reasoning that is needed to do well on those inventories.  Other assessments and PER based activities have also been based on noticing particular areas where students perform much worse than physicists.

\subsection{Use of deliberate practice in upper-division courses}

Applying this theoretical framework of deliberate practice to design instruction for upper-division courses requires one to look at the educational goals of the course.  Different courses have different goals and hence different aspects of expertise are relevant.  ``Core'' upper-division courses, such as the 3rd year course in electricity and magnetism that has been studied extensively at Colorado, involve expertise that is relevant to a set of courses and broader aspects of the undergraduate physics program.  As such, it is somewhat messy and time consuming (but very important) to establish the educational goals and objectives, taking into account the variety of perspectives and stakeholders in the outcomes of such courses \cite{PhysRevSTPER.8.020107}.  

There is another general type of upper-division course, the specialty course that goes very deeply into a specific area of physics. This is often but not always taken as an elective, and generally taught only by those faculty who are active researchers in the respective area of physics.  The goals for this type of course are essentially to have students think about and use the material much as a practicing physicist in the field (such as the instructor) does. While much the same could be said about most physics courses, defining what this means in operational terms is considerably simpler for this type of course.  For most experimental physicists in a research field, the way the material is used at the advanced undergrad level, if thoroughly mastered, is what is typically what is called upon on a day-to-day basis.  Since the instructors for such courses are frequently exactly those physicists, they can operationalize the goals and define the material needed to be covered simply by reflecting on what they and others in their field regularly do in their research activities.  What knowledge (including concepts, formalism, and instruments), procedures, approaches, and metacognitive skills do they use in solving ``routine'' (in a research context) problems?  As usually taught, such a course consists of using lectures and textbook and/or scientific literature to present the students with the knowledge and formalisms used, and there are substantial homework sets. Lectures are often organized to help students model expert thinking, with homework then being the practice of the thinking demonstrated in class.  Students, particularly the most successful ones, typically work collaboratively on the homework.   

As noted above, a fundamental component of the expertise in question, and so an essential educational goal, is to have students be able to carry out complex quantitative calculations to both understand and predict behavior in relevant contexts.  Such problems involve the blending of conceptual and quantitative skills.  This introduces novel instructional challenges, as previous PER work had not demonstrated instructional approaches that produced improvements in this type of mastery, relative to traditional lecture instruction.  Here, we demonstrate how to provide students with the necessary practice and feedback in the requisite skills, and we measure the substantial improvements in mastery that result.

From a deliberate practice perspective, rather than have class time spent having the students watch expert thinking being modelled, it is more effective to divide that expert-thinking into sections and tasks of appropriate size and difficulty, and then have the students carry out those tasks during class.  Usually those tasks are selected around particular topics and introduce knowledge as part of the exercise, but there are some elements of expertise that transcend specific topics and so are made a piece of many or all activities. An example of the latter type is checking solutions through dimensional analysis.  The second way class time can be more educationally effective from a deliberate practice perspective is to optimize the feedback.  Through the process of working directly with others in a structured way, students are guaranteed the many benefits of collaborative learning including getting immediate feedback on their ideas from others and avoiding extended time ``being stuck''.  The instructor can monitor how all the students are doing and provide very targeted and timely feedback to guide their thinking.  This is the type of formative feedback that has been seen in cognitive psychology and science classroom studies to be important for learning. \cite{hattie_power_2007,black_assessment_1998,Ambrose_2010}

In this paper we discuss the specific example of changing a specialty course in advanced optics in this manner.  This course was fairly conventional, covering such standard topics as Fourier optics, interferometers, diffraction, etc. using standard advanced sophisticated mathematical treatments for this material.  One slightly unusual feature was that the class (with an enrolment of typically 50-70 students) was composed of three distinct cohorts: ``honors'' students ($\approx 25$\% of class) who are planning to go to graduate school in physics and have completed a curriculum aimed at that goal, ``majors'' students ($\approx 25$\% of class) who are getting a physics degree but following a less rigorous program and typically not intending to go to graduate school, and Òengineering physicsÓ (ENPH) students ($\approx 50$\% of class), who are completing an engineering degree, but with a physics focus.  The ENPH cohort is the largest and, due to the program's entrance requirements, the most highly qualified. The course is required the ENPH major and for the ``honors'' students.

We used this deliberate practice framework to transform our lecture notes from previous years into active learning activities. Below we discuss in detail the transformation process for these activities and overall design guidelines, the implementation of active learning activities in the classroom environment, and our measured outcomes. The deliberate practice design framework used to transform this course could be applied in a very similar way to any such advanced specialty course.  

Another key question that we investigate here and that has been partially addressed in prior work concerns the sustainability of learning outcome improvements from course transformation and the successful continuation of a course transformation when developed materials are transferred to a new instructor. \cite{chasteen_but_2012,PhysRevSTPER.9.023102} Can the same outcomes be expected when the course materials for an active learning class are created by one instructor and then transferred to another instructor who is familiar with the topic but has never taught the course in an active learning environment? Evidence from previous studies suggests that learning outcomes are independent of class composition, but may be more dependent on the instructor but evidence is very limited.

\section{Formulation of In-class Activities}  \label{sec:form}

\subsection{Process and Specific Examples}

Following the work of Chasteen \emph{et al.}, our first step to creating class activities from course notes was to identify in each lecture the principal learning goals \cite{chasteen2011thoughtful}.  These goals are usually quite diverse and ranged from helping the students to develop a conceptual picture of a physical phenomenon to guiding the students through an important derivation or calculation. Whereas in a standard lecture presentation, learning goals are typically addressed by a series of factual statements followed by a discussion of those statements, the general philosophy for designing in-class activities is to instead frame the material and address the learning goals in a series of questions. The second step is to compose a question/task or calculation whose response would require drawing upon or formulating a working understanding of these learning goals. In our activities, sometimes a single question would involve multiple learning goals while in other cases a single learning goal would be addressed by a series of shorter questions ordered in increasing conceptual difficulty or computational complexity.

Figure~\ref{afig:optical-resonator-derivation_act} (in Appendix A) shows an in-class activity involving the derivation for the field inside an optical resonator while Fig. \ref{afig:optical-resonator-derivation_lect} shows the corresponding lecture (that were actually given to students in a traditional class) from which the activity was derived.  The activity is meant to lead the students through the derivation by breaking the steps down into a series of questions.  Questions about the interpretation of various steps are also included so that students can simultaneously build their physical intuition about the solution as they work on the derivation. In this example the derivation is simple enough (in terms of mathematical operations/algebra) that an effective in-class activity can be constructed to step through it. In contrast,  Fig.~\ref{afig:Huygens-Fresnel} (in Appendix A) presents an activity on the Huygens-Fresnel principle and the corresponding diffraction  integral expression relating the electric field in the $(x,y)$-plane at two different $z$ values. As this derivation is much more mathematically sophisticated (and thus would require a disproportionately large amount of class time), rather than step through the derivation, the final expression is provided and students are asked to explain the physical meaning of the terms in the expression and to relate them to a statement of the Huygens-Fresnel principle.  As before, also included are lecture notes given to students when used in a class taught in a lecture format.  In both cases, the activities follow very closely the formulation of the topics from the lecture notes, but the delivery is inverted.  For example, Huygens-Fresnel principle, instead of the instructor explaining or stating the Huygens-Fresnel principle and then discussing it, the students are asked to recall or reconstruct from their reading the statement of the principle and then explain its relation to a mathematical expression. In essence, we ask the students to mirror how a practicing physicist would connect mathematical models to the physics under study. 

\subsection{General Design Considerations}

We summarize our primary design guidelines in Table \ref{table:design} that we have empirically found result in well-received and effective in-class activities.
\begingroup
\linespread{1.2}
\begin{table*}[h]
\caption{Design considerations used for this course transformation}
\centering

\begin{tabular}{l}
\toprule
\hline
$\bullet$ Each activity was motivated either verbally by the instructor or in the written preamble.\\
$\bullet$ Mathematical models were tied to phenomena, and ``real-life'' examples were employed.\\
$\bullet$ Prior course material was included wherever possible to maximize continuity.\\
$\bullet$ Questions were ordered from least to most difficult to optimize engagement.\\
$\bullet$ Activities were designed so that continuous work did not exceed 15 minutes and at least \\ two \indent class feedback sessions could occur over the full class period.\\
$\bullet$ Activities included some work that was judged to be just beyond what the students \\ were capable of, so as to optimally prepare them for the feedback period.\\
$\bullet$ Long/elaborate algebraic manipulations in activities were minimized so students could focus \\ on underlying physical phenomena. \\
$\bullet$ ``Bonus'' questions were included to challenge the most advanced students. \\

\bottomrule

\end{tabular}

\label{table:design}
\end{table*}
\endgroup
Our experience has convinced us that that two important design considerations for optimal student engagement are the duration and the structure of the activity.  Specifically, multiple activities each lasting between 5 and 15 minutes (usually closer to the latter) are optimal; the students stay more fully engaged and ``resynchronizing'' the class after 15 minutes with new activities prevents the students from dispersing too broadly in their progress on completed activities.  Starting activities with simple introductory questions followed by more complex questions provides a natural way to maintain high class engagement in cases where there is a wide range of prior knowledge or experience.  This structure prevents less prepared students from struggling without any success while still allowing more advanced students to find intellectual challenges in the material.    

Such structuring is illustrated in top half of  Fig.~\ref{fig:appliedproblems} in Appendix A which presents an activity where the students analyze the use of an interferometer to measure the acceleration due to gravity.  The activity is structured in four questions which begin as basic and qualitative and become progressively more complex and quantitative, with the final ``bonus'' question extending the activity well beyond the scope of the original problem. We label the latter as ``bonus'' meaning that we don't expect the majority of students to complete this particular question, but it keeps those students engaged who complete the activity more quickly than the majority of the class. 

This example along with a second in the bottom half of Fig.~\ref{fig:appliedproblems} on lunar ranging also illustrate the design principle of enhancing student engagement by tying what they were learning in class to ``real'' world applications. We witnessed a noticeable improvement in student interest/engagement using these types of problems.

In addition, we found an open-ended ÒworksheetÓ format worked better than multiple-choice type questions in this course.  Students are able to work at their own individual group pace (before being synchronized as discussed below) on activities, as well as allowing more complex questions than allowed by multiple choice answers. Moreover, as one overall goal of this advanced course was to train students as practicing experimental physicists, it is more appropriate to present activities in such a more realistic open-ended format. 

Finally, we discovered that difficulties associated with understanding proper mathematical notation required special treatment to avoid the students spending an excessive amount of time on struggling with definitions and labels. In this case a brief mini-lecture was given at the start of class, providing a clean example to illustrate the notation; this is particularly important if multiple texts are used with different notation conventions. An example of notation that we would treat in this way in our course is the complex amplitude representation of electro-magnetic waves.

\section{Classroom Implementation}  \label{sec:class}

The progression through a typical class is shown in Table \ref{table:class} along with the corresponding roles and actions for both the students and instructors/TAs.  Each action of students and instructors is discussed in detail below in the corresponding sections. 

\begingroup
\linespread{0.8}

\begin{table*}
\ra{1.3}
\caption{Progression through sequential stages of a typical class period. Each action of students and instructors is described in detail within text.}
\centering

\begin{tabular}{  l  l l l }

\toprule
\hline
{\bf Actions}       & \bf{Time (min)} & \bf{ Students}                                                                                 & \bf {Instructors/TA(s)}                                                                                          \\ \hline
 
Preparation     & |                             & \begin{tabular}[c]{@{}l@{}}Complete targeted reading \\ and pre-class online quizzes\end{tabular}  & Formulate/review activities                                                      \\ \hline
Introduction           & 2-3                            & Listen/ask questions                                                                                   & Introduce goals of day                                                                                  \\ \hline
Activity        & 10-15                          & Group work on activities                                                                               & \begin{tabular}[c]{@{}l@{}}Circulate in classroom, answer \\questions and assess students\end{tabular} \\ \hline
 Feedback & 5-15                          & \begin{tabular}[c]{@{}l@{}}Listen/ask questions, \\provide solutions and \\ reasoning when called on\end{tabular}                                                                     & \begin{tabular}[c]{@{}l@{}} Facilitate whole class \\ discussion, provide feedback \\to class \end{tabular}                                                                    \\ \hline
Activity        &        \multicolumn{1}{l}{ \vdots}                       &           \multicolumn{1}{l}{ \vdots}                                                                           &                            \multicolumn{1}{l}{ \vdots}                                                                               \\ \hline
 Feedback &      \multicolumn{1}{l}{ \vdots}                     &          \multicolumn{1}{l}{ \vdots}                                                                             &                            \multicolumn{1}{l}{ \vdots}    \\ \hline
\ldots {\it repeat as needed} \ldots   &     \multicolumn{1}{l}{ \vdots}                               &                  \multicolumn{1}{l}{ \vdots}                                                                              &                           \multicolumn{1}{l}{ \vdots}          \\ \hline
Conclusion        & 2-3                             & Hand in in-class work                                                                    & Wrap up    \\ 
\bottomrule
\end{tabular}
\label{table:class}
\end{table*}
\endgroup

\subsection{Preparation}
In addition to the preparation of in-class activities discussed in Section \ref{sec:form}, a pre-class meeting between the instructor and TAs to review the activities and discuss any anticipated questions/difficulties is necessary. During this meeting we also estimate the time needed for each activity as a rough guide to refer to during the class.

For students, prior to each class, they are assigned reading to prepare them for the in-class activity.  Completion of the reading is important for effective participation in the in-class activities.  We have found three guidelines helpful to encourage and facilitate completion of this reading.\cite{Heiner_reading_2014} First, the pre-reading is targeted and specific to only the upcoming class.  Rather than assigning something generic like ``read Chapter 2'', an example assignment would be: 

\begin{quote}
Read Chapter 2 \S3 pp 49-50, Chapter 2 \S4 pp 40-55. {\it Make sure you understand the relationship between Eqn. 2.3 and 2.4. If you are having troubling understanding Eqn. 2.7, that is okay but make sure you understand everything before it.} 

\end{quote}	

For a 90-minute class we assign at most 12-15 pages of reading; the average is 8-9 pages. We use targeted reading assignments of this length to encourage students to complete the reading and enable them to use their out-of-class time most effectively.   Second, online quizzes on the reading before each class provide an additional, direct incentive to read.  These quizzes consist of one or two questions designed to take only a few minutes for the students if they have completed the reading. Collectively, these quizzes counted for 7.5\% of a students total mark (almost as much as homework) and are due a few hours before class begins. For reference the total mark distribution in the course is given in Appendix C. Third, we found it important to remind students, throughout the term, of the importance of reading, particularly when answering studentsÕ individual questions during in-class activities (see below).

\subsection{Introduction}

At the start of each class the instructor first gives a quick (2-3 minute) summary of the upcoming topics and activities of the day. Students are also encouraged to ask any questions on their pre-class reading before the instructor moves on to the first in-class activity. While important to respond to this engagement, we were careful not to go into lecture mode and spend too much time answering such questions.  Whenever possible we defer any possible long explanation to the upcoming activities.

\subsection{Activity}

The first set of activities is displayed (usually projected) to the class and students are allowed to self-organize into groups numbering between two and five members and begin work on the activities. To facilitate the student group work, the classroom had movable tables and chairs. During this period, students are not allowed to use any books or computers, thus requiring them to work from memory and to engage in critical thinking to construct the information they need to answer the questions. In other words, they practice how physicists and engineers analyze new concepts/techniques from the beginning. These conditions also promoted engagement and discussion of students within their group. We did, however, allow them to use any personal notes on the reading.  Although they work collaboratively, each student is required to record his/her own solutions to the activities, and these solutions are turned in at the end of each class and marked for participation (counting for 7.5\% of the total mark).

During the students' group activity work, the instructor and TA(s) circulate through the room assessing progress and answering questions.  The instructor/TAs also assess written work and ask groups questions to gauge their understanding and progress.  If there is little cooperative dialog between group members, the instructor/TA first pose a question to one group member and then asks other members to comment or provide an alternate answer (thereby initiating dialog).  We have found the ratio of students to instructors/TAs can be as high as 35:1 before running into difficulty answering most of the student questions and performing an adequate class-wide assessment of progress during a 10-15 minute activity.  We note that not every group is visited by either the instructor or TA during a given activity for this assessment. Compared to the pre-reformed course, this transformed course did not require substantially more TA time. Specifically, it required only one additional TA to help with the in-class activities and to post the pre-reading quizzes. With a class size of 35 students or less, no additional TA help is necessary for running the in-class sessions compared to the traditionally taught course.  To operate at this high ratio, we have found the following guidelines helpful:

\begin{itemize}
\item{the instructors/TA need to continuously move around with a goal of spending only a few minutes at most with each student group/question;}
\item{if there is a common misunderstanding or question shared by several groups, the instructor temporarily interrupts work on the activities to address the class as a whole on this issue. This interruption usually lasts only a few minutes at most (it does not turn into an extended lecture, rather it is just clarifying a point, question or concept);}
\item{if the instructor/TA notices that one group has successfully addressed a misconception while a nearby group is still struggling with same issue, then the instructor asks the first group to serve as the ``expert'' to provide guidance to the second group, thereby freeing the instructor/TA to move onto to other groups. This strategy also leverages peer-wise instructional benefits \cite{Ambrose_2010} as well as renews the engagement of more advanced students who might have already finished more of the activities.}
\end{itemize}

\subsection{Feedback}

At intervals normally dictated by natural breaks in activity topics (but at most 15 min) work by the students on the in-class activities is suspended and the instructor provides feedback by discussing solutions for the preceding activities. The students are well primed for this feedback by their previous work. This procedure also serves to keep the class progressing by resynchronizing all the students before moving on to the next set of activities. We have found the duration for activities can be quite variable and the time to begin feedback is best decided on the spot based on class progress.  We normally wait until a majority of students have worked through the basic concepts and have spent some time considering, if not working on, the more in-depth questions. Each example activity in Figs.~\ref{afig:optical-resonator-derivation_act}-\ref{fig:appliedproblems} consist of one activity block. 

When delivering feedback, the instructor works through a prepared solution of the activities, while simultaneously engaging the students for input, comments, and questions on the solutions as they are presented.  If we note a particularly clear or alternate solution written by a student, we use it in place of  (or as a supplement to) the instructorÕs solutions by projecting the studentÕs solution via a document camera so all can see it. Normally, bonus activity solutions are provided to students on-line but not discussed during class. After class, the solutions to the activities are posted on-line for students to use as a reference.    

\subsection{Conclusion}

The activity-feedback sequence is repeated throughout class, ending always with feedback. We almost never complete all the prepared activities, as the intent is to push students to their limits. The unused activities are usually applied to the next class unless a new topic is scheduled. If the last feedback session is artificially shortened due to time constraints (occurring in approximately 25\% of classes) we specifically ask students to finish looking over solutions outside of class. However, if unfinished activities were particularly important for the following material we assign the activity as homework due at the beginning of the following class. In our case, this situation occurred two or three times a term.   

\section{Deliberate Practice Efficiencies}  \label{sec:eff}

A common expectation is that the total material that can be covered using this type of instruction will be significantly less than by lecturing. The pace of the lecture is typically set by the speed at which an instructor writes out the notes and then discusses them, while the pace of the transformed course is set by the time the instructor allows the students to spend on each set of activities and the time spent in the feedback period. As this deliberate practice instruction was directly developed from a set of well-honed lecture notes that had been used in previous iterations of this course, we can make a quantitative comparison in the amount of material covered between the two approaches. We found that about 10\% fewer hours were used in lecture to present the material covered in the transformed course and observers of this instructor's (DJ) lectures judged the pace to be average for upper-division physics courses.   As a result, in the transformed course we continued to cover fundamental topics (plane waves, paraxial waves, Gaussian beams, interferometers, Fourier optics, optical resonators, and polarization) while dropping some specialized material (covered in the lecture course) that we felt was nonessential and included wave guides, electro-optic and acoustic-optic devices.

To cover this much material in the transformed course we take advantage of two sources of inherent efficiency:

\begin{itemize}
\item{class time can be spent almost entirely on the difficult concepts in both activity and feedback stages, spending very little time on material students already know or learn easily;}

\item{time-intensive mathematical derivations/manipulations can be replaced by presenting the calculation as a whole and asking students to identify the critical steps and the underlying physics. The activity in Fig.~\ref{afig:Huygens-Fresnel} is an excellent example of this attribute.}

\end{itemize}

To maximize coverage we also found it is important to create activities with clearly defined problem statements that students can quickly understand, have targeted pre-class readings so students are able to quickly engage with the activity, and regularly resynchronize the entire class to ensure some groups do not fall too far behind. 

\section{Measured Outcomes}  \label{sec:outcomes}

\subsection{Evaluation Methods}

The student populations measured are shown in Table \ref{table:pop}. Prior to transformation, this course was taught using a traditional lecture format for three years by a professor without any prior active learning experience (DJ, nominated by students in one of these years for a University-wide teaching award for this course).  The students from the last of these years form the control group (CG). There are two experimental groups. The first group (EG1) consists of students in the first term that the course was transformed and taught by the original professor (DJ) with help from an STLF.  The second group (EG2) is made up of students taking the transformed version of this course four years later taught by a second professor (KM) who did have prior active-learning teaching experience but only normal TA support (no STLF).  The transformed course materials for EG2 were largely the same as that from the original transformation for EG1.

\begingroup
\begin{table}[h]

\caption{Populations of the Control Group (lecture-based course), Experimental Group 1 (transformed course taught by original professor, DJ) and Experimental Group 2 (transformed course taught by second professor, KM) broken down by specific student programs. The physics student populations (both Honors and Majors) vary from year to year, while historically the ENPH are nearly constant in both number and quality; the reason for the population increase in EG2 was a change in frequency the course was offered (from twice a year to once a year) between EG1 and EG2.}
\ra{1.3}
\centering
\begin{tabular}{ l  l  l  l l }

\toprule
\hline
{\bf Group}       & \bf{ENPH~~} & \bf{Honors~~}                                                                                 & \bf {Majors~~}         & \bf{Tot.}                                                                                 \\ 
\hline

Control Group (CG)     		      & 25                          &13    & 14         & 52                                             \\ 
Exp. Group 1 (EG1)~~           & 23                          & 3      & 5     & 31                                                                            \\ 
Exp. Group 2 (EG2)~~           & 53                          & 14   &  14 & 81         \\ 
\bottomrule
\end{tabular}

\label{table:pop}
\end{table}
\endgroup

We evaluate learning outcomes using a set of common final exam problems (taken closed book) that test major learning goals of the course.  Two examples of these problems are shown in Fig.~\ref{fig:Exam} in Appendix B. These problems are specifically designed to evaluate critical thinking skills rather than formulaic recollection by requiring students to apply skills and concepts in contexts different from those in which they had previously seen them used. 
 
In all cases the problems were almost identical (isomorphic) for the control and experiment groups.  For all of the years in this study, the final exams were not handed back to students and no exam problems were used twice, other than these used in this study.  We believe that these problems were not in circulation and the time ordering of the control and experimental groups is not of concern.  For this study, a single grader using the same rubric marked all exam solutions of students from all three groups for each exam problem. 

\subsection{Results and Discussion}

\begin{figure}[h!]
\centering
\resizebox{0.49\textwidth}{!}{%
  \includegraphics{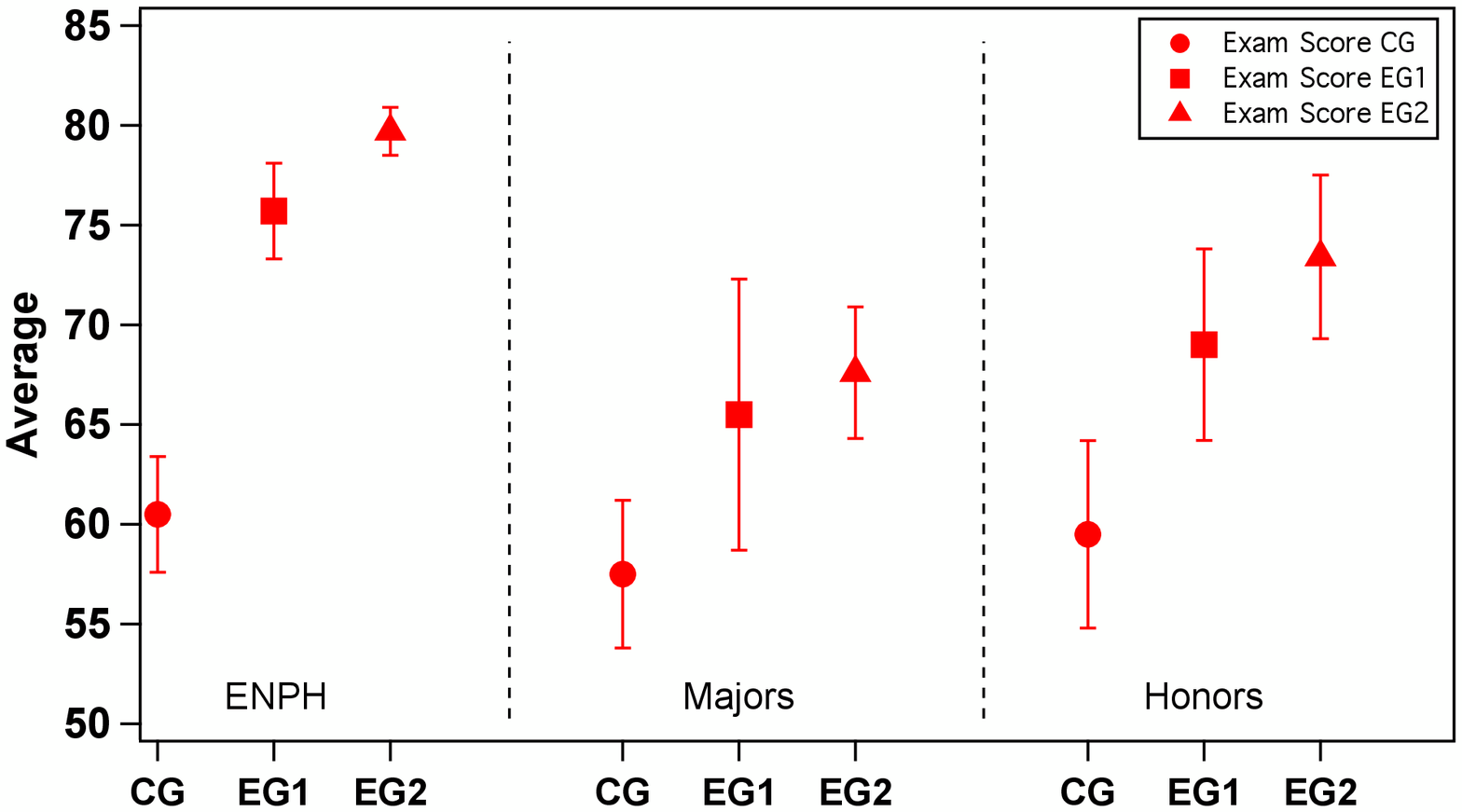}}
\caption{Normalized exam score average of the engineering physics (ENPH), physics majors and physics honors students in the control (lecture-based) group, CG and experiment groups (transformed), EG1 and EG2. Error bars shown on the exam scores are the standard error.}

\label{fig:results}       
\end{figure}

Results of normalized exams scores are listed in Table \ref{tab:results} and plotted in Fig.~\ref{fig:results}. There is a significant improvement of the exam scores for the transformed courses (EG1 and EG2) across all three student cohorts. The learning gains persist when the course is taught four years later by a different instructor.  This demonstrates that the gains are robust and independent of instructor 
and that the course materials are easily transferable.  We note that a faculty member who has never used active learning would need to familiarize himself or herself with the format and structure of the class, but we do not believe that there are any skills not already possessed by a faculty member required to run such a course.  
\begingroup
\squeezetable
\begin{table}[h]
\caption{Normalized scores on test problems of the control group (lecture-based) and experiment groups (transformed) for the three types of students. Standard error is shown in $()$.  $\dagger$Three different topics/problems were used in these exam scores, while the remainder had two topics/problems tested; in all cases nearly identical problems were used for each topic. Hedges' $g$ effect size for EG1 and EG2 (relative to CG) are also shown for the three groups. $\ddagger$The ENPH CG score is made up from three problems, while the ENPH EG2 score is a subset of two of those problems. Accordingly, $g$ of ENPH EG2 is calculated using the corresponding two problem normalized score for ENPH CG which was 62.8 (4.4).}
\ra{1.3}
\begin{tabular}{ l  l  l  l  l  l  l }

\toprule
\hline
                & \multicolumn{3}{c}{\textbf{Exam Scores (\% correct)}}                             &\phantom{abc}                 & \multicolumn{2}{c}{\textbf{Effect Size $g$}} \\ 
                \cmidrule{2-4}  \cmidrule{6-7}
\textbf{Group} & \textbf{CG}    & \multicolumn{1}{l}{\textbf{EG1}} & \multicolumn{1}{l}{\textbf{EG2}} && \textbf{EG1}        & \textbf{EG2}        \\ \cmidrule{2-4}  \cmidrule{6-7}
ENPH           & 60.5$\dagger$ (2.9)~                   & 75.7$\dagger$ (2.4)~ & 79.7 (1.2)~ && 1.16                & 1.19$\ddagger$                \\ 
Honors~~ & 57.5 (3.7)~                   & 65.5 (6.9)~ & 67.6 (3.3)~ && 0.54                & 0.75                \\ 
Majors~~ & 59.5 (4.7)~                   & 69.1 (4.8)~ & 73.4 (4.1)~ && 0.57                & 0.84                \\ 
\bottomrule
\end{tabular}

\label{tab:results}
\end{table}
\endgroup

To further quantify the effect of the course transformation upon learning outcomes we also calculate Hedges' $g$ effect size of exam scores using a pooled standard deviation 
\cite{borenstein_introduction_2009} with the results summarized in Table \ref{tab:results}. In this context, effect size quantitatively measures the effect on learning outcomes of the transformation to active learning (EG1 and EG2) relative to lecture-based learning (CG). Mathematically, effect size is equal to the difference between the means divided by the standard deviation and Hedges' $g$ is a refinement that includes a factor for the finite sample size. An effect size of $>0.8$ is considered a large/significant effect. \cite{cohen_statistical_1988} For the ENPH population, the original transformation produced an effect ($g=1.16$) and ($g=1.19$) for the second instructor, significantly beyond the ``large'' effect designation.  It is of interest to note both of these effect sizes are  $\approx 60\%$ larger than the mean effect size reported \cite{Freeman12052014} in a meta-analysis of active learning studies across STEM courses.  This result is a important indication that peer work in an active learning classroom is at least as effective for advanced students studying advanced material as it is for lower division students in introductory courses.

In our study the much lower standard error and higher effect size for the ENPH students compared to other students is due largely to two main factors. First, this study had a significantly higher number of ENPH students compared to Physics Honors and Majors students (see Table \ref{table:pop}). Second, at UBC, the high demand for, and accompanying high selectivity, of the ENPH program means the year-to-year student population is exceptionally stable in terms of number, class average, and overall abilities.  As noted elsewhere\cite{Freeman12052014}, observed gains in active learning environments do not show a significant dependence on student's background or program of study; therefore, we expect our ENPH results (with its better statistics and stable population) to represent a closer measure of the effect size in our transformation to active learning. 

This analysis points to another significant result of our study: a quantitative confirmation that improved learning outcomes obtained by an initial transformation to an active learning environment are robust. After four years a second instructor was able to use the material without significant modifications and obtained comparable results.

To confirm the stability and consistency of the ENPH students across CG, EG1 and EG2 we tracked each student's performance through two other traditionally taught upper-division physics courses (PHYS 301, a first term E\&M course and PHYS 350, a classical mechanics course). While a majority ($>$65\%) of students in CG and EG1 took both of these courses in the same term (Fall 2007) and a similar percentage in EG2 took them in Fall 2010, due to individual timelines the remainder took them in other years.  We thus defined the following metric which is the average number of standard deviations each group is away from the course average to evaluate if one particular group is stronger or weaker than another,

\begin{equation}
\bar{D}=\frac{1}{N}\sum_{i=1}^ND_i =\frac{1}{N}\sum_{i=1}^N\left[\frac{s_i-\bar{s_i}}{\sigma_i}\right]
\label{eqn:D}
\end{equation}

where $s_i$ is the final mark of the $i^{th}$ student, and $\bar{s}_i$ ($\sigma_i$) is the course average (standard deviation) of the course the $i^{th}$ student took.  The total number of students in PHYS 301 and 350 ranged from 50 to 145, thus providing good statistics. As shown in Table \ref{tab:tracking}, the metric of CG was {\it higher} than the average by $0.25 \sigma$ from the average, while for EG1 it was only $0.08 \sigma$ above the average and for EG2 it was $ 0.18\sigma$ below the average. The control and experimental ENPH groups went through the same curriculum and pre-requisite courses before taking our course, and, based on higher averages in two of these courses (PHYS 301 and 350, the former of which is a prerequisite to this optics class), it is highly unlikely that CG had a weaker background or were composed of weaker students prior to the beginning of our course and therefore performed worse on our assessments. These results in fact show the CG group to be at least as strong if not stronger than either experimental groups in these other upper-division physics courses.

\begingroup
\begin{table}[h]
\caption{The number of standard deviations each group's final grade average was from the class average ($\bar{D}$ as defined in Eqn.~\ref{eqn:D}) when these students took two other non--transformed upper-division physics courses. The standard error of $\bar{D}$ is shown in parentheses. Also of note is that the standard deviation of the $D_i$ values are very close to 1 in units of $\sigma$, the standard deviation of the final grade distribution.  This indicates that these subsets of students (the control and experimental groups) were distributed in much the same way and thus not atypical of the overall class distribution itself.}
\ra{1.3}
\begin{tabular}{ l  l  l  l  l  l   }

\toprule
\hline
                & \multicolumn{2}{c}{\textbf{E\&M}}                             &\phantom{abc}                 & \multicolumn{2}{c}{\textbf{Classical Mechanics}} \\ 
                \cmidrule{2-3}  \cmidrule{5-6}

\textbf{Group~~~~} & $\bar{ \textbf{{D}} }$   &  \textbf{$\sigma_D$}  && $\bar{ \textbf{{D}} }$         &  \textbf{$\sigma_D$}         \\ \cmidrule{2-3}  \cmidrule{5-6}

CG           & 0.22 (0.04)~~                   & 0.95   && 0.27  (0.04)~~                & 0.96               \\ 
EG1~~ & 0.07 (0.04)~~                    & 0.99 && 0.08 (0.03)~~                & 0.86                  \\ 
EG2~~ & -0.15 (0.02)~~                    & 0.97  && -0.22 (0.02)~~                & 0.91                  \\ 
\bottomrule
\end{tabular}

\label{tab:tracking}
\end{table}
\endgroup

A remaining possible concern with the CG based on a single term is that the instructor (DJ) could have had an uncharacteristically ``bad'' term teaching.  An examination of DJ's student evaluations prior to and including the CG term indicate this possibility to be highly unlikely. In these evaluations completed two years prior to the CG, the result of the question: ``Overall the instructor was an effective teacher'' (on a Likert scale, with 5 strongly agree, etc.) was 4.5 (0.10), 4.4 (0.17), and 4.4 (0.17) with the last number representing the CG term.  Moreover, this was the fourth year for DJ to teach this course and while we didn't do quantitative comparisons with previous years (when he taught it traditionally) the overall student performance was similar. The performance of CG in other upper-division physics classes shown in Table \ref{tab:tracking} supports these observations.

Although implemented in a specialized upper-division course for experimental modern optics, the transformation process and accompanying learning improvement should be applicable to most all upper-division physics courses with an established set of lecture notes.  It should apply equally to experimental and theoretical topics.

\begingroup
\linespread{1.0}
\begin{table*}[b]
\caption{Results from two attitude surveys in the first transformed year with average response. The standard error is shown in $()$.}
\centering
\begin{tabular}{ l  l  l  l  l  }
\toprule
\hline
{\textbf{Survey Question}} & \multicolumn{2}{l}{\textbf{Likert scale choices}} &  \multicolumn{2}{l}{\textbf{Student}} \\ 
 & (1) top & (5) bottom  &\textbf{responses} \\ 

  \multicolumn{2}{l}{\textbf{Mid-term}}  & & &\\ \hline
I am learning a great deal in this course & strongly agree~~ & strongly disagree~~ & 1.68 (0.12) \\ 
My group is working together & very well & awfully & 1.63 (0.09) \\  

 \multicolumn{2}{l}{\textbf{End-of-term}} & & & \\ \hline

I am learning a great deal in this course & strongly agree & strongly disagree & 1.45 (0.16) \\ 
For my learning, the in-class activities were~~ & very useful & worthless & 1.23 (0.13) \\  
For my learning, the pre-class reading was & very useful & worthless & 1.59 (0.18) \\  
\bottomrule
\end{tabular}
\label{table:survey}
\end{table*}
\endgroup
We do not have measures of time spent studying for either the lecture or transformed classes so it is possible the observed improvements could have resulted in part from increased out of class work by students in EG1 and EG2. We reduced the homework for the transformed class by approximately 25\% compared to the lecture class in response to initial student feedback regarding overall workload of the transformed class; the workload likely (and understandably) increased because students are required to read prior to each lecture and were encouraged to review the in-class activities and the posted solutions. This amount of homework reduction was meant to equalize the workload and was based on the assumption that no pre-class reading was being done in the untransformed course.  We believe it achieved the desired workload balance and is supported by informal student feedback.

%


To assess student attitudes towards the class, we conducted two anonymous surveys in the first transformed year.  The first survey was done in the middle of the term and the second was conducted at the end of the term. The results are shown in Table V and indicated that the students saw the value of the learning activities.

\section{Conclusion}  \label{sec:conclud}

In this work, we report on the process and outcomes of transforming a specialized upper-division physics course using active learning pedagogies and we observe large improvements in learning for all student cohorts as measured by isomorphic (nearly identical) final exam problems when both the control and experiment groups are taught by the same instructor.  We also observe that these learning improvements are sustained when the course was taught by a different instructor using the same in-class activity materials.  Our process for converting from lecture notes to questions for in-class learning activities, allows 90\% of material that could be covered compared to the traditional lecture format.  Finally, we present data on student attitudes and the results indicate a high degree of satisfaction and acceptance of the active learning format.  We believe that a similar process could be employed for any speciality upper-division physics course.

\begin{acknowledgments}
The authors acknowledge significant contributions from Louis Deslauriers and Ellen Schelew. Our work was supported by the University of British Columbia through the CWSEI. 
\end{acknowledgments}

\bibliography{PER-3}
\newpage
\appendix
\section{Examples of In-class Activities}

\begin{figure}[htbp]
\centering
\includegraphics[scale=0.75]{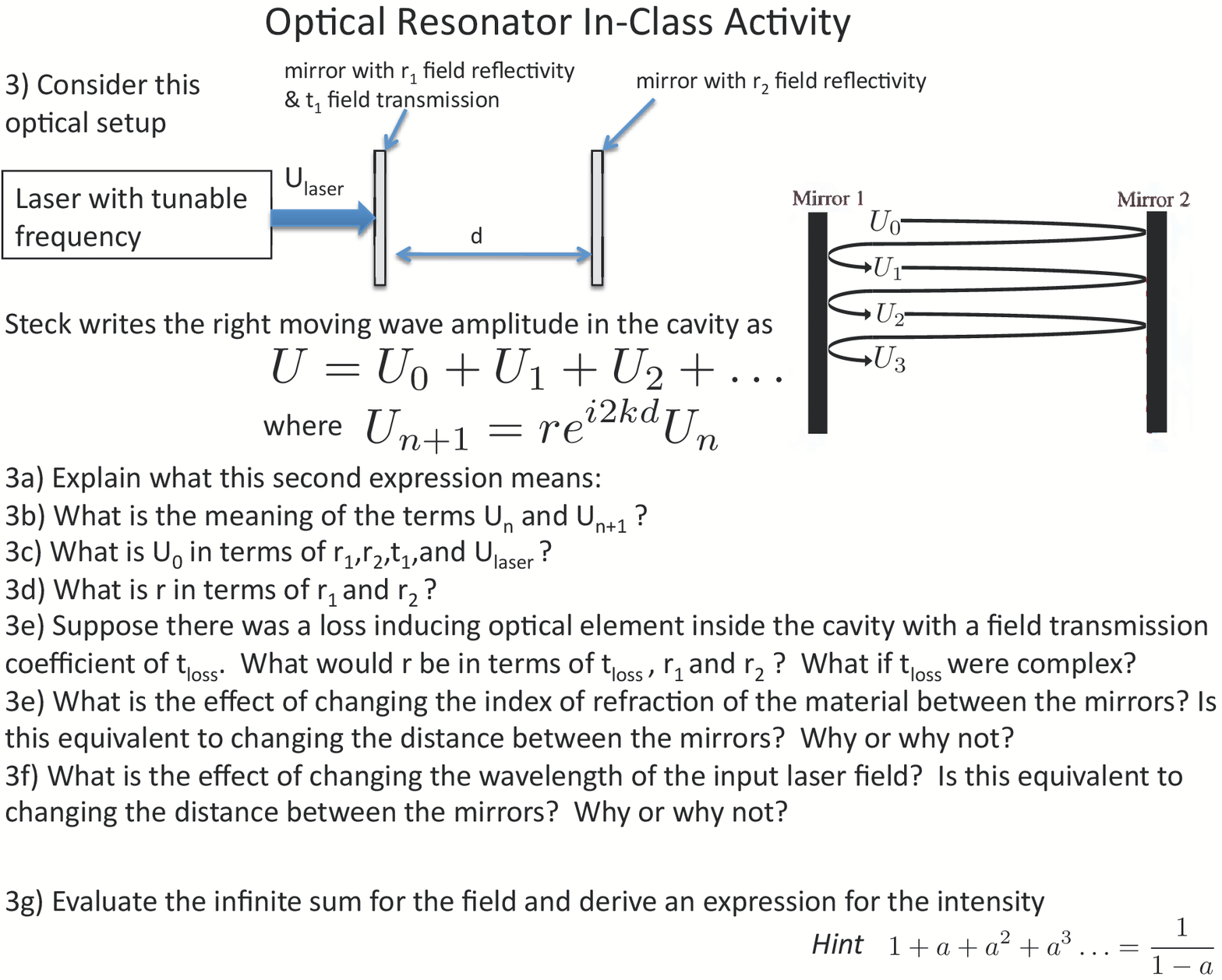}
\caption{An in-class derivation for the field inside an optical resonator.}
\label{afig:optical-resonator-derivation_act} 
\end{figure}

\begin{figure}[htbp]
\centering
\includegraphics[scale=0.75]{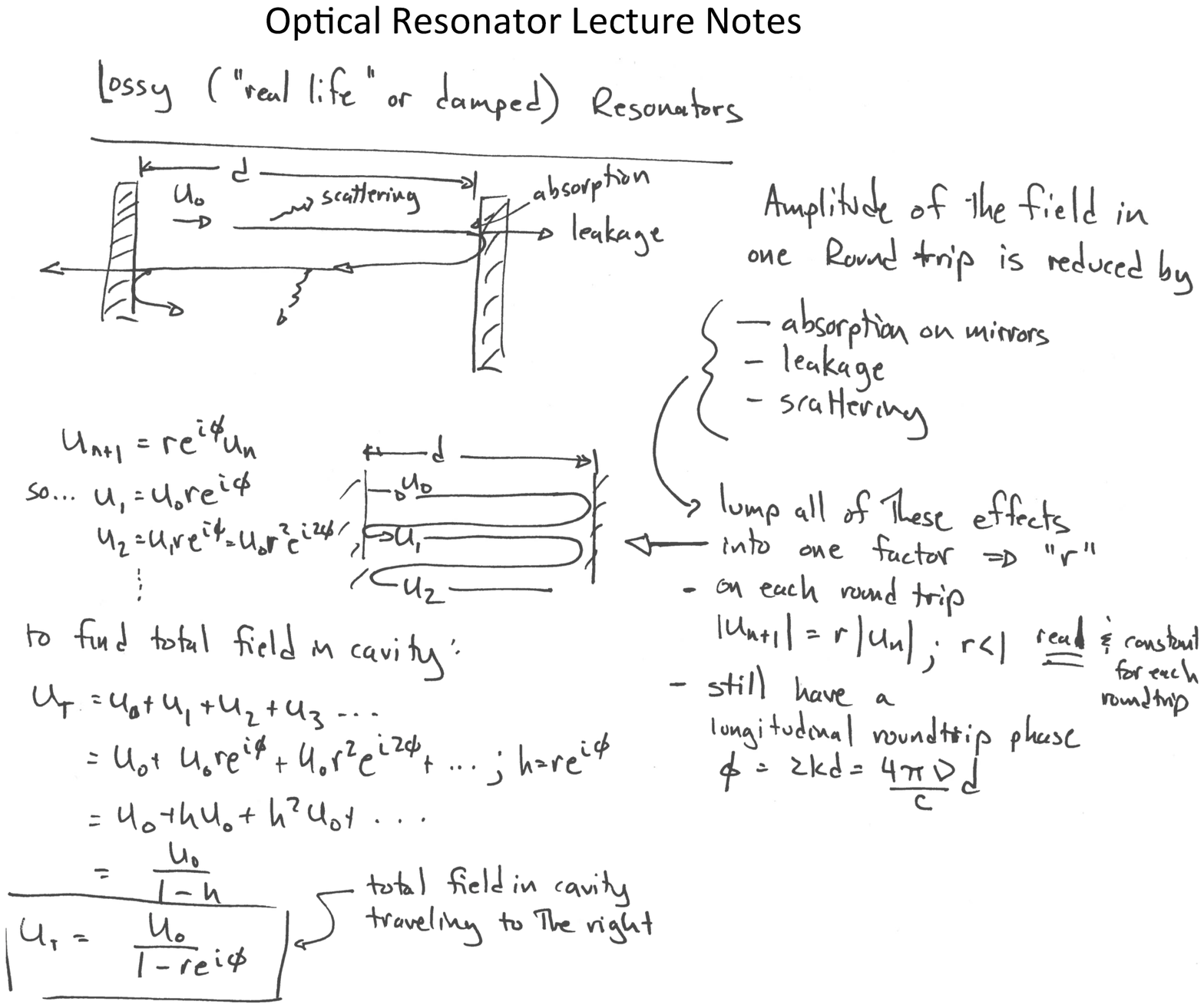}
\caption{The corresponding lecture notes for the in-class activities in shown in Fig. \ref{afig:optical-resonator-derivation_act}}
\label{afig:optical-resonator-derivation_lect} 
\end{figure}

\begin{figure}[htbp]
\centering
\includegraphics[scale=0.80]{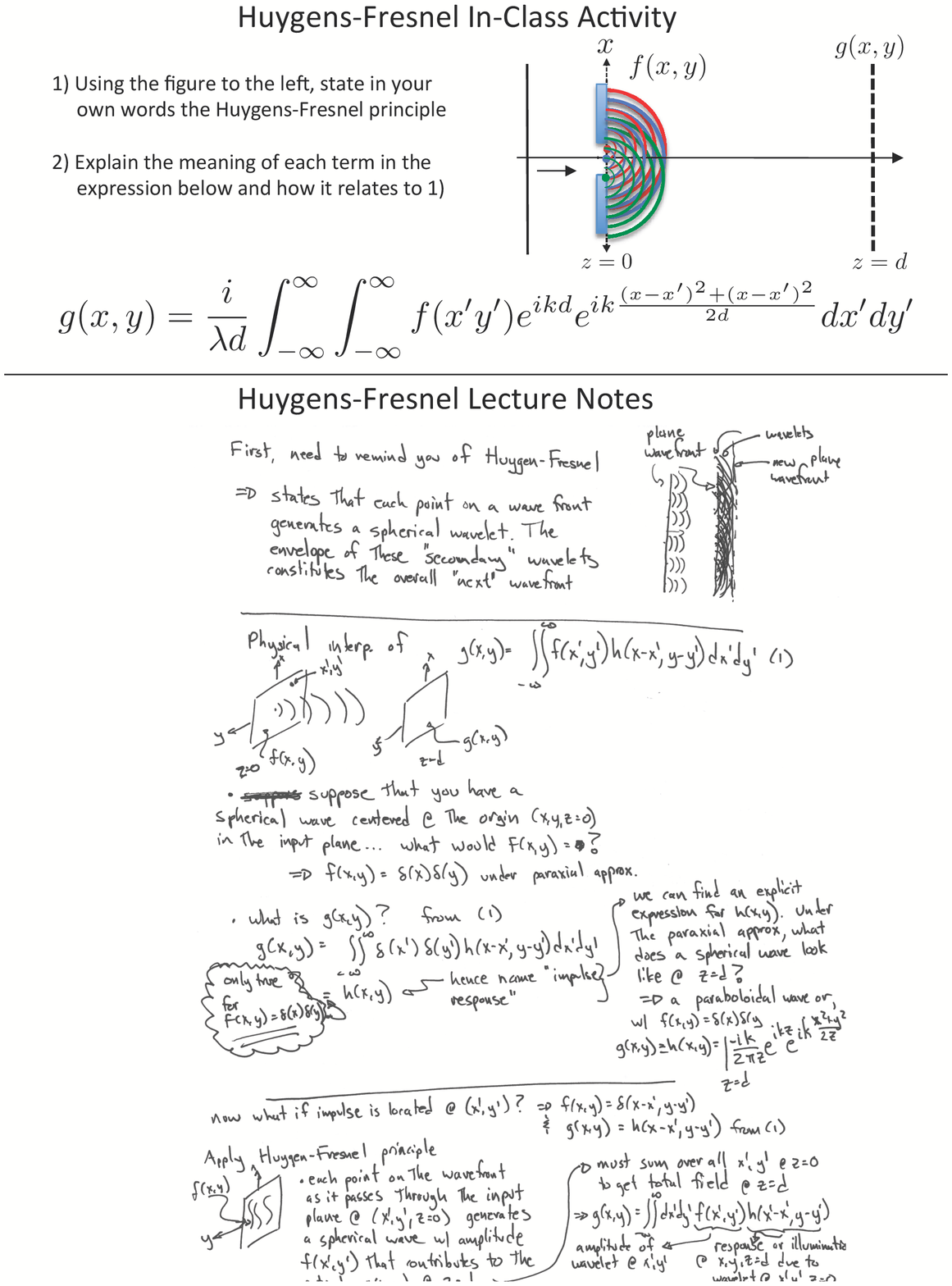}
\caption{An in-class activity on the Huygens-Fresnel principle and the corresponding lecture notes.}
\label{afig:Huygens-Fresnel}
\end{figure}

\begin{figure}[htbp]
\centering
\includegraphics[scale=0.80]{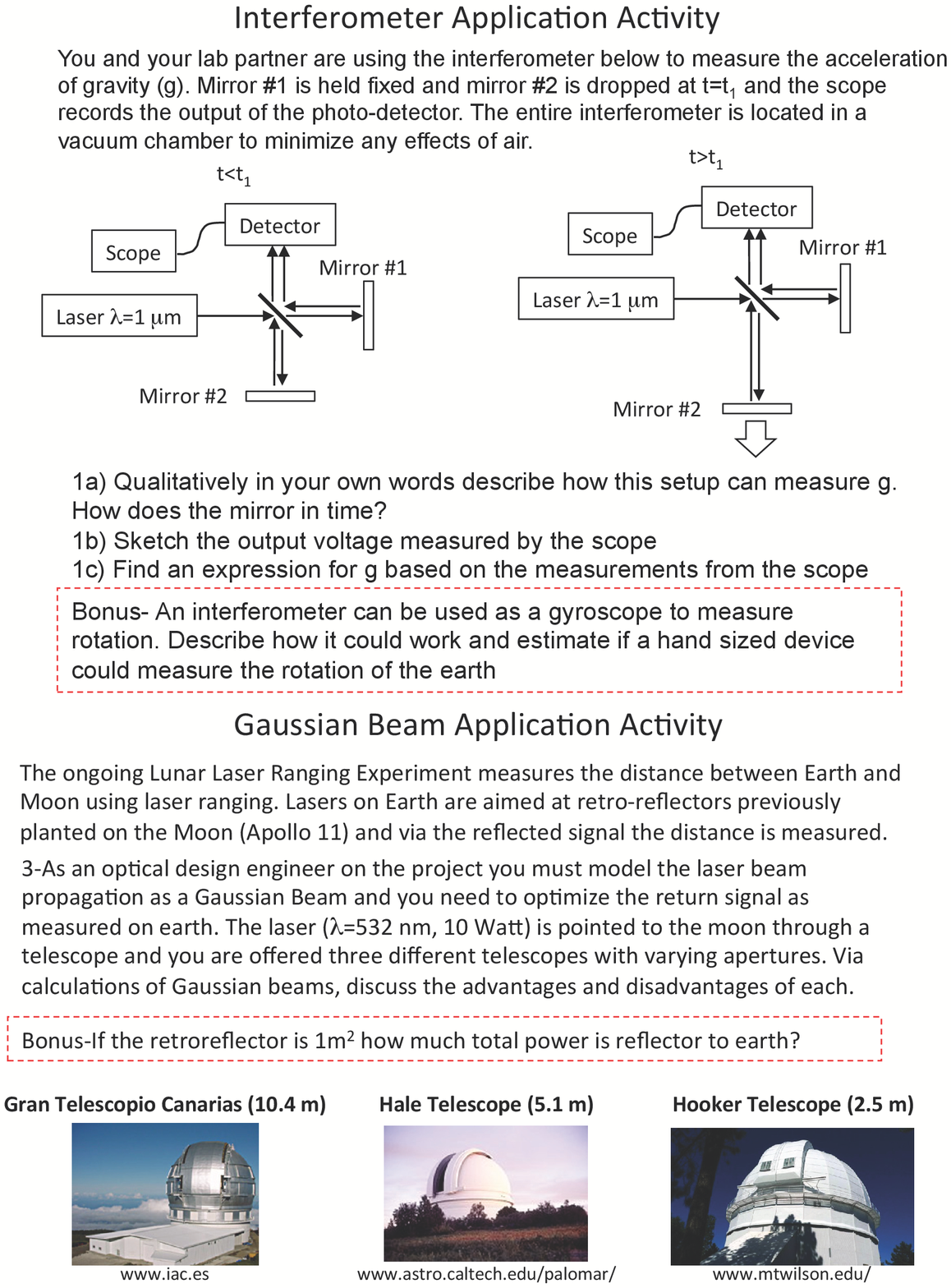}
\caption{In-class activities on the applications of optics to measure the acceleration due to gravity and for lunar ranging.}
\label{fig:appliedproblems}
\end{figure}

\clearpage
\newpage
\section{Examples of Exam Problems} 

\begin{figure}[htbp]
\centering
\includegraphics[scale=0.70]{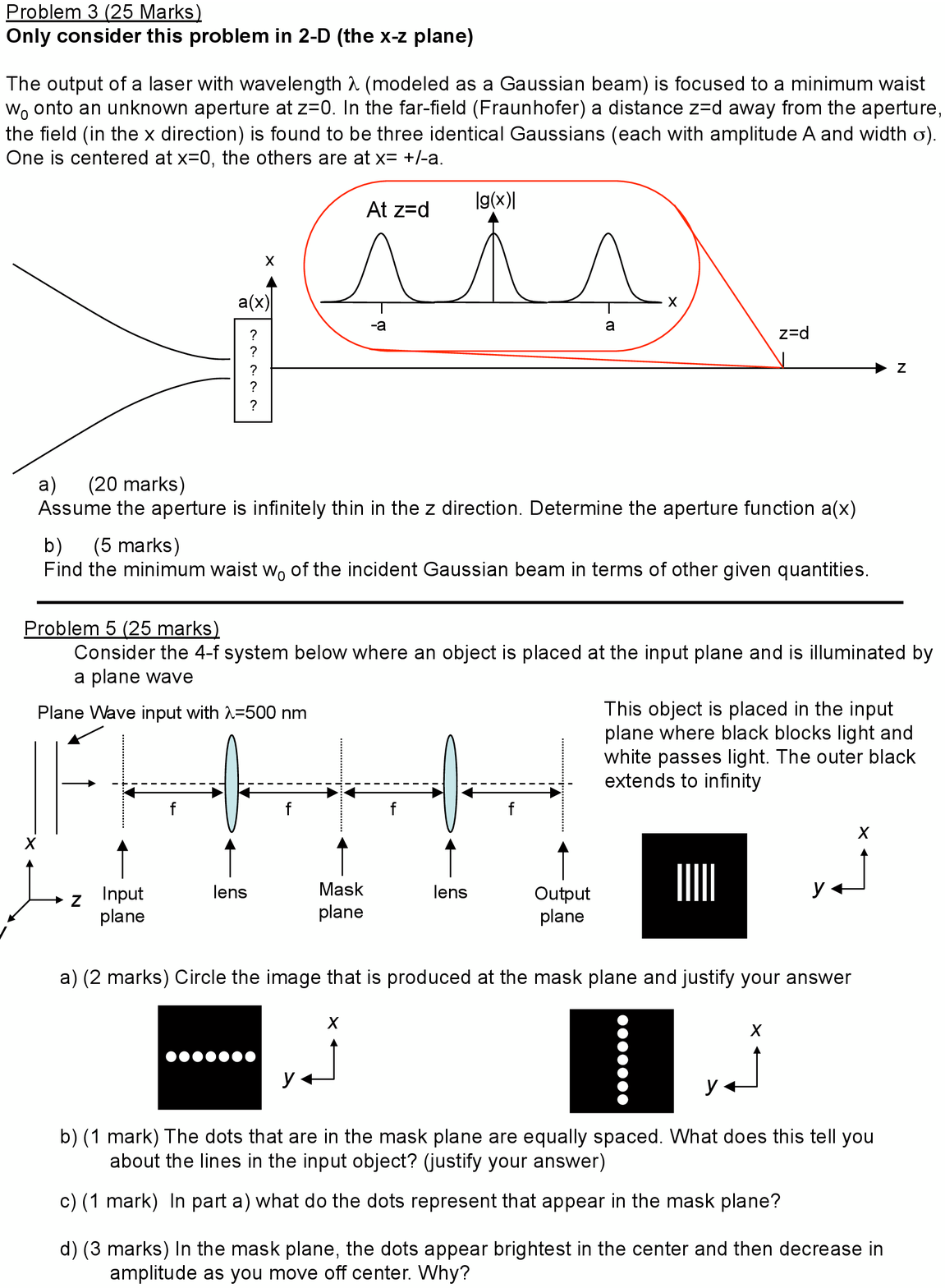}
\caption{Examples of exam problems covering subjects of Gaussian beams, diffraction, and Fourier Optics.}
\label{fig:Exam}
\end{figure}

\begin{figure}[htbp]
\centering
\includegraphics[scale=0.85]{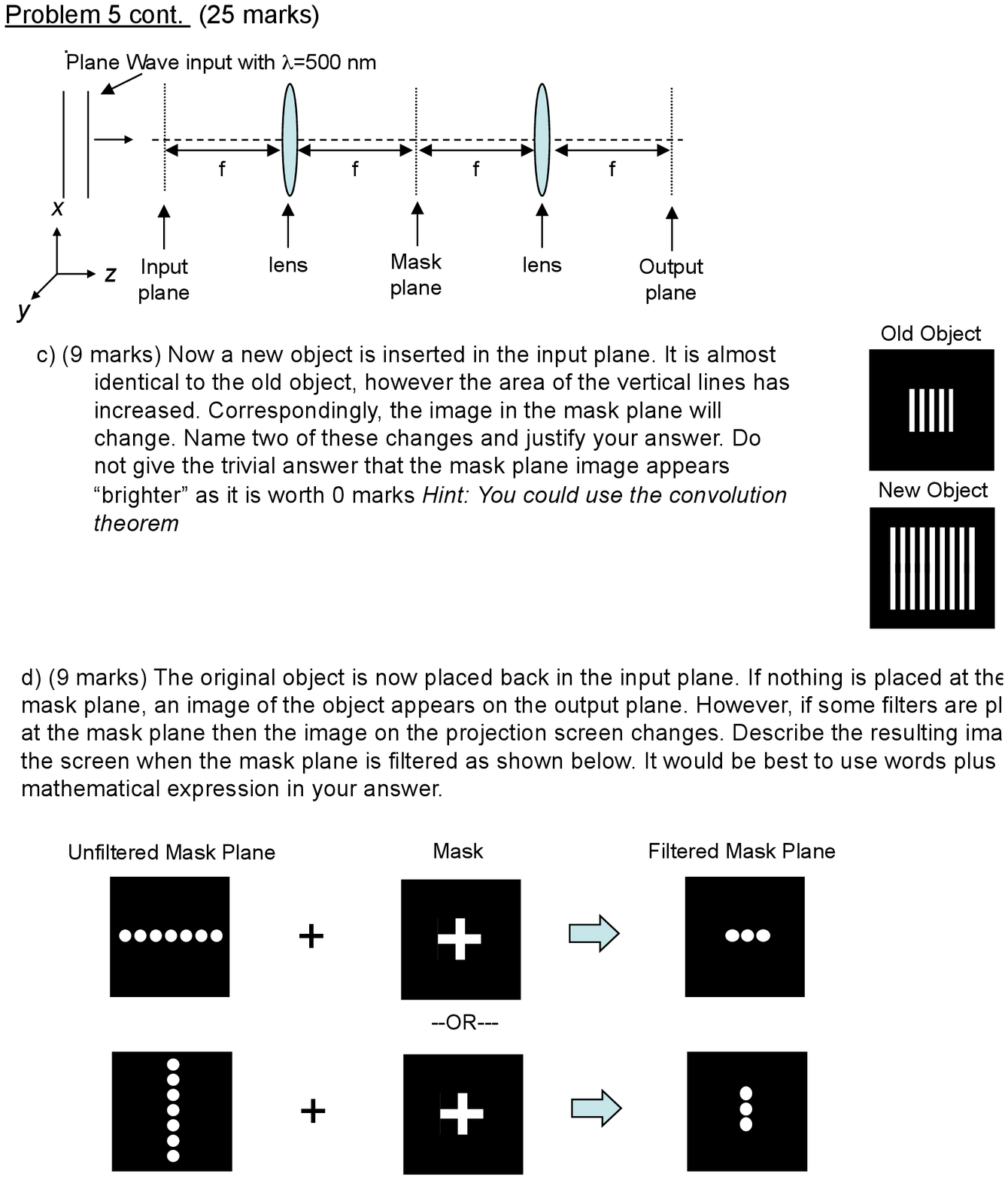}
\caption{Continued example of exam problem covering Fourier Optics.}
\label{fig:Exam2}
\end{figure}

\clearpage
\newpage

\section{Mark Distribution} 
\begingroup
\begin{table}[!h]

\caption{Mark distribution used in course. Note the relatively heavy weight (and thus incentive) of the pre-reading quizzes and in-class activities.}
\ra{1.3}
\centering
\begin{tabular}{ l  l   }

\toprule
\hline
{\bf Item}       & \bf{Percentage}  \\ 
\hline
Pre-reading Quizzes & 7.5\\
In-class Activities/Participation~~ & 7.5\\
Homework & 10\\
Laboratory & 20\\
Midterm & 20\\
Final & 35\\

\bottomrule
\end{tabular}

\label{table:mark}
\end{table}
\endgroup

\end{document}